\pdfoutput=1
\documentclass[aps,pra,reprint,superscriptaddress,longbibliography]{revtex4-2}

\usepackage{amsmath,amsfonts,amssymb,bm,upgreek,mathrsfs,mathcomp}
\usepackage{graphicx}
\usepackage{siunitx}
\usepackage[dvipsnames]{xcolor}
\usepackage{hyperref}
\usepackage{cleveref}
\hypersetup{
    colorlinks=true,
    linkcolor=blue,
    citecolor=blue,
    urlcolor=blue,
    hypertexnames=false
}
\crefname{figure}{Fig.}{Figs.}
\crefname{equation}{Eq.}{Eqs.}
\newcommand{\ucite}[1]{\cite{#1}}
\begin{document}

\title{Fast and Coherent Transfer of Atomic Qubits in Optical Tweezers using Fiber Array Architecture}

\author{Jia-Chao Wang}
\affiliation{Division of Precision Measurement Physics, Wuhan Institute of Physics and Mathematics, Innovation Academy for Precision Measurement Science and Technology, Chinese Academy of Sciences, Wuhan, China}
\affiliation{School of Physical Sciences, University of Chinese Academy of Sciences, Beijing, China}

\author{Zai-Zheng Zhang}
\affiliation{Division of Precision Measurement Physics, Wuhan Institute of Physics and Mathematics, Innovation Academy for Precision Measurement Science and Technology, Chinese Academy of Sciences, Wuhan, China}
\affiliation{School of Physical Sciences, University of Chinese Academy of Sciences, Beijing, China}

\author{Xiao Li}
\email{lixiao@apm.ac.cn}
\affiliation{Division of Precision Measurement Physics, Wuhan Institute of Physics and Mathematics, Innovation Academy for Precision Measurement Science and Technology, Chinese Academy of Sciences, Wuhan, China}

\author{Guang-Wei Wang}
\affiliation{Division of Precision Measurement Physics, Wuhan Institute of Physics and Mathematics, Innovation Academy for Precision Measurement Science and Technology, Chinese Academy of Sciences, Wuhan, China}
\affiliation{School of Physical Sciences, University of Chinese Academy of Sciences, Beijing, China}

\author{Xiao-Dong He}
\affiliation{Division of Precision Measurement Physics, Wuhan Institute of Physics and Mathematics, Innovation Academy for Precision Measurement Science and Technology, Chinese Academy of Sciences, Wuhan, China}

\author{Min Liu}
\affiliation{Division of Precision Measurement Physics, Wuhan Institute of Physics and Mathematics, Innovation Academy for Precision Measurement Science and Technology, Chinese Academy of Sciences, Wuhan, China}

\author{Peng Xu}
\email{xupeng@apm.ac.cn}
\affiliation{Division of Precision Measurement Physics, Wuhan Institute of Physics and Mathematics, Innovation Academy for Precision Measurement Science and Technology, Chinese Academy of Sciences, Wuhan, China}
\affiliation{Wuhan Institute of Quantum Technology, Wuhan, China}

\begin{abstract}
Programmable neutral-atom arrays offer a promising route toward scalable quantum computing, where coherent qubit transfer enables non-local connectivity and reduces resource overhead. However, transfer speed and motional heating remain key bottlenecks for fast and deep quantum circuits. Here, we employ a fiber array neutral-atom quantum computing architecture with site-resolved control of trap depths to realize smooth amplitude exchange between static and moving traps, thereby enabling fast and coherent qubit transfer with ultralow motional heating. With a 10 $\mu$s in situ transfer between static and moving traps, we obtain a per-cycle heating rate of 0.156(9) $\mu$K, sustain over 500 cycles with negligible atom loss, and achieve a quantum state fidelity of 0.99992(5) per cycle. For inter-site transfer between two separated static traps, the operation takes 120 $\mu$s with 0.783(17) $\mu$K heating per transfer, and remains negligible atom loss for up to 100 repeated cycles with a fidelity of 0.9998(1) per transfer. Furthermore, through experimental studies of parallel transfer, we establish a model that elucidates the relationship between array inhomogeneity and the transfer heating rate. This fast, low-heating coherent transfer capability provides a practical route for improving both speed and fidelity in atom-shuttling based quantum computing.

\end{abstract}

\maketitle

\section{Introduction}
	
    Neutral-atom arrays trapped in optical tweezers have emerged as a promising platform for quantum computation and quantum simulation\ucite{Bluvstein2026, Anand2024, Reichardt2025, XiaolingWu2021, DongQiMa2025,Li2025}. A distinctive advantage of these systems is their ability to provide on-demand, non-local connectivity between qubits, which is particularly beneficial for logical-qubit encoding and logical gate operations\ucite{Zhou2025,Bluvstein2024, Bluvstein2022, Baspin2022, Xu2024,D'Angelis2020,Baspin2022connectivity,Saffman2025}. Typically, realizing such non-local connectivity requires coherently transferring qubits between tweezer sites across the array, thereby bringing selected qubits into a desired configuration for interaction or addressing\ucite{Manetsch2025,Hwang2023,Klostermann2022,Lengwenus2010,Pagano2024,Yang2016,Bluvstein2022}. To date, this capability has been demonstrated using several approaches, including operating under magic-intensity trapping conditions to suppress differential light shifts\ucite{Yang2016} , employing adiabatic protocols to minimize motional heating\ucite{Bluvstein2022}, and leveraging AI-assisted optimization of tweezer depths and positions to improve transfer fidelity and robustness\ucite{Manetsch2025}.

However, current schemes lack independent, site-resolved control of the depth of each static trap (S-trap), which prevents smoothly ramping an individual S-trap depth to zero. Under this constraint, during extraction with a moving trap (M-trap), the atom experiences a residual potential from the S-trap, introducing an additional time-dependent perturbation. To suppress the resulting motional heating, in situ transfer is typically carried out on a relatively long timescale, often requiring hundreds of microseconds.  Moreover, owing to accumulated heating, repeated transfers over just tens of cycles lead to noticeable atom loss\ucite{Reichardt2025}, which constrains performance in high-speed, large-depth quantum circuits.

In this work, we leverage a fiber-array architecture with site-resolved, independent control of static-trap depths\ucite{Li2025} to realize a fast coherent transfer scheme based on smooth amplitude exchange between static and moving traps\ucite{Beugnon2007}. Using this approach, we demonstrate in situ coherent transfer in 10 $\mu$s between static and moving traps with a per-cycle heating rate of 0.156(9) $\mu$K, sustaining 500 lossless cycles. Building on the transfer above and combining it with atom transport, we further demonstrate low-heating inter-site transfer between two separated static traps. Finally, we perform parallel inter-site transfer and establish a model that quantifies the dependence of the transfer heating rate on array inhomogeneity, informing the optimization of coherent transfer in large-scale arrays.

\begin{figure}[t]
    \centering
    \includegraphics[width=0.48\textwidth]{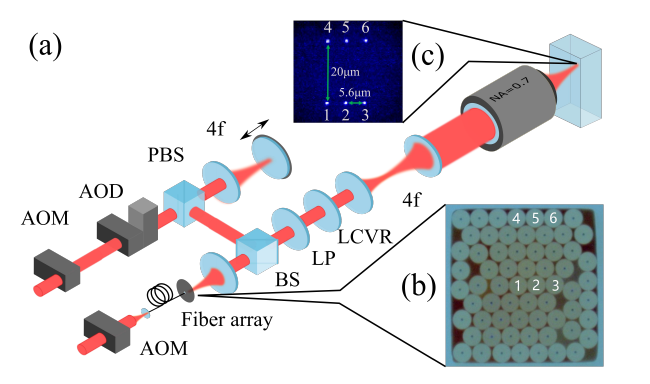}
    \caption{Schematic of the experimental setup for coherent transfer.
(a) The M-trap laser beam first passes through the acousto-optic modulator (AOM) and acousto-optic deflector (AOD). Then it is combined with the S-trap laser beam using a beam splitter (BS). The combined beams subsequently pass through a linear polarizer (LP) and a liquid crystal variable retarder (LCVR), followed by a 4f beam-expanding system and an objective, before finally entering the vacuum chamber. (b) The cross-section of the fiber array. Six traps were selected from the array for our experiment. (c) Fluorescence image of six single atoms trapped in selected traps.}
    \label{fig:fig1}
\end{figure}
\section{Experimental Setup}	

The experimental setup of the fiber-array-based atomic quantum computer has been described in detail in our previous work\ucite{Li2025}. Each fiber serves as the interface for trapping and addressing an individual $^{87}\text{Rb}$ atom, simultaneously enabling scalability and independent control. In this work, we utilize fiber-array-based optical traps as S-Traps, selecting six of them, each with independently controllable depth, and introduce an M-Trap for qubit transport within the array (see \cref{fig:fig1}).  After emerging from the AOD, the M-Trap beam first passes through a reflective 4f relay system with $1:1$ magnification, and then combined with the S-Trap beam using a beam splitter ($T:R = 9:1$).

For fast, low-heating in situ coherent transfer, precise alignment between the S-trap and M-trap is required. Radial alignment is achieved by scanning the AOD frequency and fitting the atom survival probability after repeated transfer cycles, while axial alignment is optimized by translating a mirror in the reflective 4$f$ relay system. The measured trap-frequency difference between the M-trap and S-trap is below 3\%. More details of the experimental setup are provided in the Supplementary Material.

\section{Fast Coherent in situ Transfer of a Single Atomic Qubit}

After aligning the S-trap and M-trap, we perform rapid trap-depth modulation to realize in situ qubit transfer. Ideally, perfect spatial overlap between the two traps maintains a constant effective trap depth during amplitude swapping, yielding zero heating regardless of the modulation waveform. In practice, however, thermal fluctuations and mechanical drifts introduce a slight spatial misalignment. Under small misalignment limit, the amplitude exchange shifts the combined trap minimum, resulting in an effective short-distance transport of the atom that induces heating. To minimize this heating, we design the depth variation so that the effective spatial motion of the trap minimum follows a shortcuts-to-adiabaticity (STA) \ucite{Ness2018, Hwang2025, Cicali2025, Couvert2008} trajectory, as detailed in the Supplementary Material. The depth variation of the M-trap is described by the following STA function:

\begin{equation}
x(t)=-20t^7 + 70t^6 - 84t^5 + 35t^4 \ (t, x \in [0,1])\tag{1}
\end{equation}

\begin{figure}[h]
    \centering
    \includegraphics[width=0.48\textwidth]{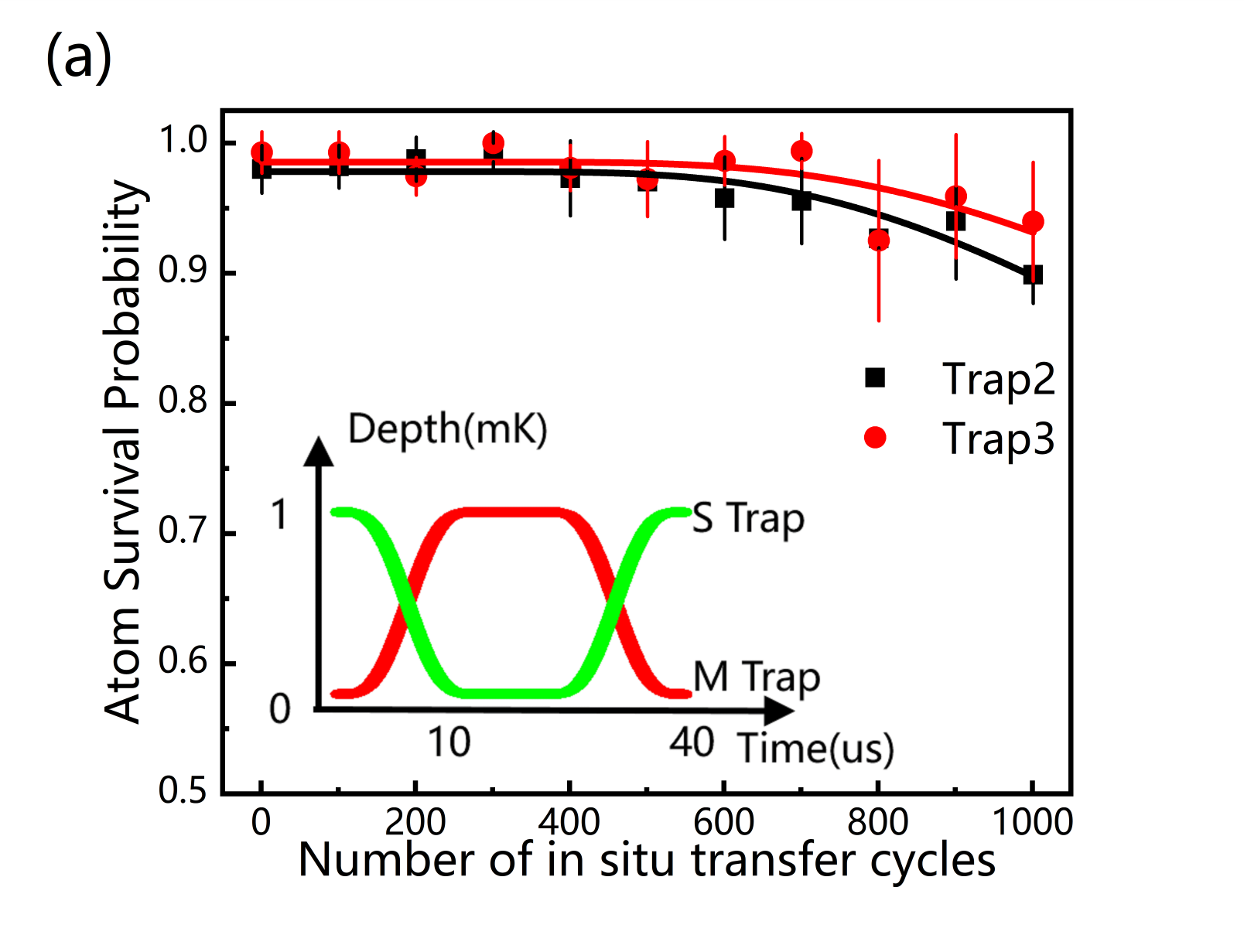}\hfill
    \includegraphics[width=0.48\textwidth]{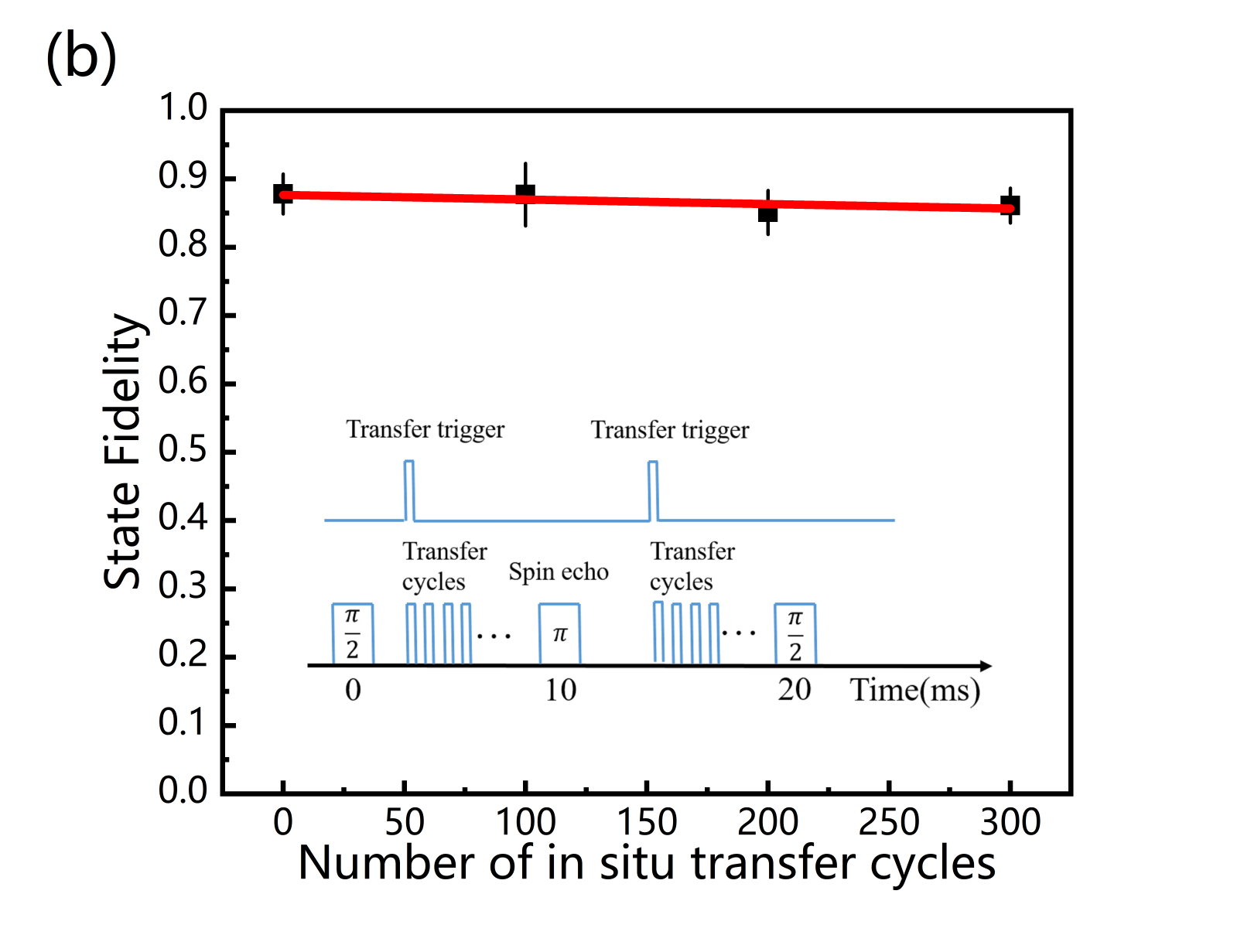}
    \caption{Fast coherent in situ transfer of a single atomic qubit.
    (a) Survival probability after different numbers of in situ transfer cycles for both directions. The inset is the transfer process in one cycle. The corresponding heating rates are extracted by the fit of the experimental data using Eq. (3).
    (b) The measured internal-state fidelity after different transfer cycles, with the one-cycle fidelity extracted by the exponential fitting function. The inset shows the microwave pulse sequence used for tomography.}
    \label{fig:2}
\end{figure}

The amplitude variation of the S-trap follows the function  $1-x(t)$ (as shown in the inset of \cref{fig:2}(a)). We maintain the effective trap depth $U_0$ of the combined S-trap and M-trap at 1 mK, with an amplitude swap time of 10 $\mu$s and a subsequent 20 $\mu$s wait time to confirm completion of  one-way transfer process. During this wait time, the initial trap remains totally off so that any untransferred atoms are lost, ensuring the measured survival strictly reflects a genuine transfer rather than the mere survival of the depth modulation. The atom survival probability has decreased from 0.98 to around 0.9 after 1000 transfer cycles. The loss is mainly due to transfer induced atom heating (see \cref{fig:2}(a)).

To analyze the heating rate, we adopt a theoretical description in which the temperature of atoms in a dipole trap follows a Boltzmann distribution\ucite{Tuchendler2008}. Here, we introduce a normalized temperature coefficient
\begin{equation}
\zeta=  ( T_0 +\Delta T\cdot n )/ U_0 \tag{2}
\end{equation}

where $T_0$ is the initial temperature, and $\Delta T$ is the heating per transfer, $n$ is the number of transfer cycles. By integrating the Boltzmann energy distribution up to the trap depth, the survival probability of the atom remaining in the dipole trap after the truncation at the finite trap depth is given by:
\begin{equation}
P(n) = P_0 -(1+1/\zeta+1/(2\cdot \zeta^2))\cdot \exp(-1/\zeta)\tag{3}
\end{equation}

where $P_0$ is  initial atom survival probability.  In our system, the initial atom temperature is $15$ $\mu \text{K}$, corresponding to $\sim 13.0\%$ of the 2D radial motional ground-state fraction. We fitted the experimental data and obtained the per transfer cycle heating rates of Trap 2 and Trap 3 as $0.165(6) $ $\mu\text{K}$ and $0.148(11)\  \mu\text{K}$, respectively, which induce a reduction of $\sim0.2\%$ in the radial motional ground-state fraction after the first transfer cycle.

Additionally, we employed quantum state tomography\ucite{Thew2002, Hofmann2004, Yu2014} to characterize the internal state fidelity of the atomic qubits during the transfer process. The two hyperfine levels of the $^{87}\text{Rb}$ clock transition are chosen as the basis states \(|0\rangle = |F=1, m_F=0\rangle\) and \(|1\rangle = |F=2, m_F=0\rangle\). By employing magic-intensity trapping and spin echo techniques, we have extended the coherence time of the atoms to over 100 ms, which is significantly longer than the coherent transfer time\ucite{Yang2016, Guo2020}. We begin by preparing the atom in \(|0\rangle\) state with a raw fidelity above 0.95, followed by a resonant \(\pi/2\) microwave pulse to prepare it in a superposition state. To suppress inhomogeneous dephasing, a spin-echo \(\pi\) pulse is inserted after \(10\ \text{ms}\). Following an interval of \(10\ \text{ms}\), a second \(\pi/2\) pulse is applied. Coherent transfers are performed symmetrically about the \(\pi\) pulse, distributed across both intervals (see inset of \cref{fig:2}(b)). After completing this 20 ms sequence, the raw state fidelity is measured at 0.88(3), which has decreased by $\sim 0.07$ compared to the initial $|0\rangle$ state preparation. This is primarily due to the photon scattering induced fidelity loss in the 830nm dipole trap with 1mK depth. After 300 transfer cycles, the state fidelity remained at 0.86(3). The single transfer cycle fidelity obtained from exponential decay fitting is 0.99992(5) (see \cref{fig:2}(b)).

\begin{figure}[!h]
    \centering
    \includegraphics[width=0.41\textwidth]{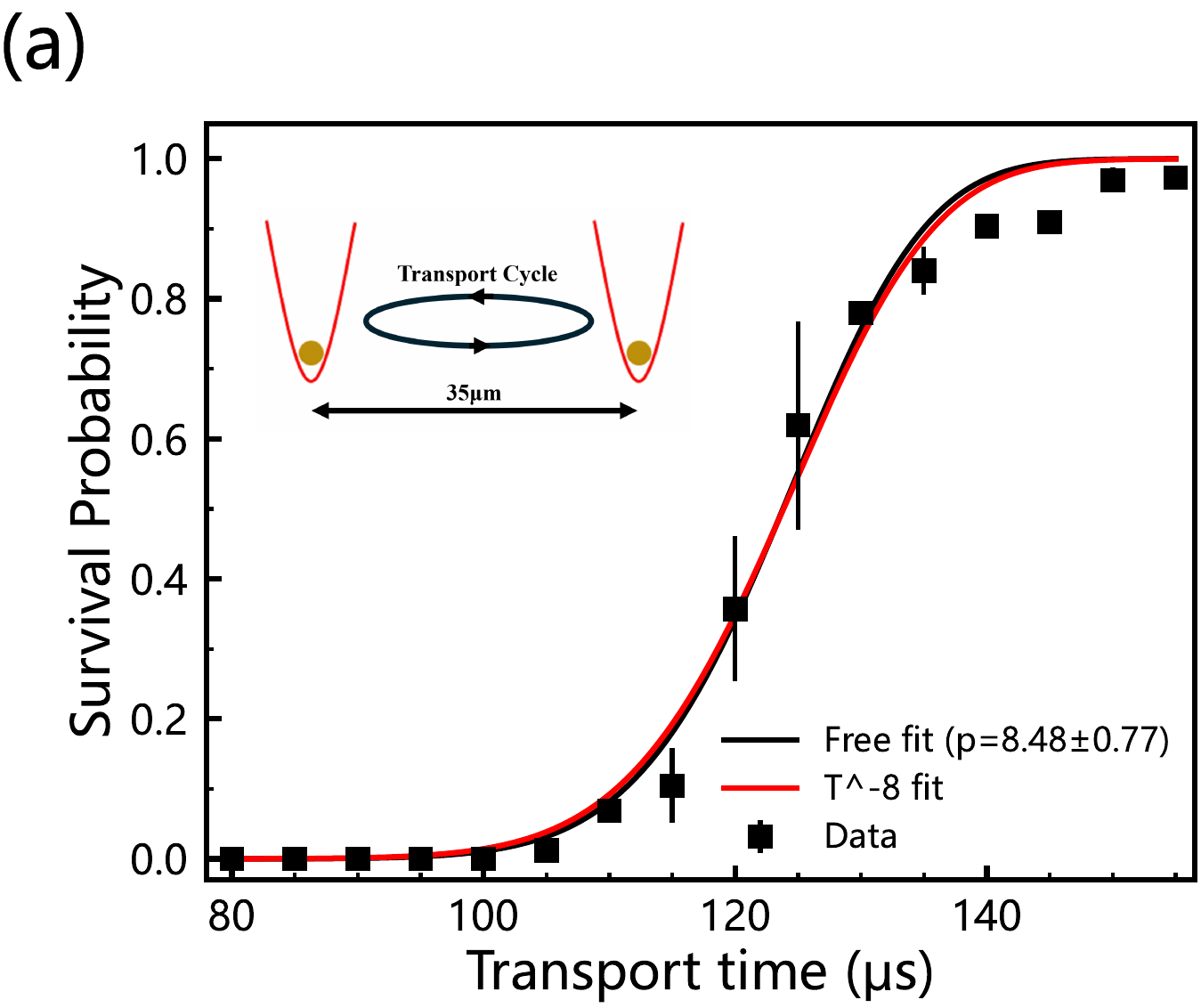}\hfill
    \includegraphics[width=0.41\textwidth]{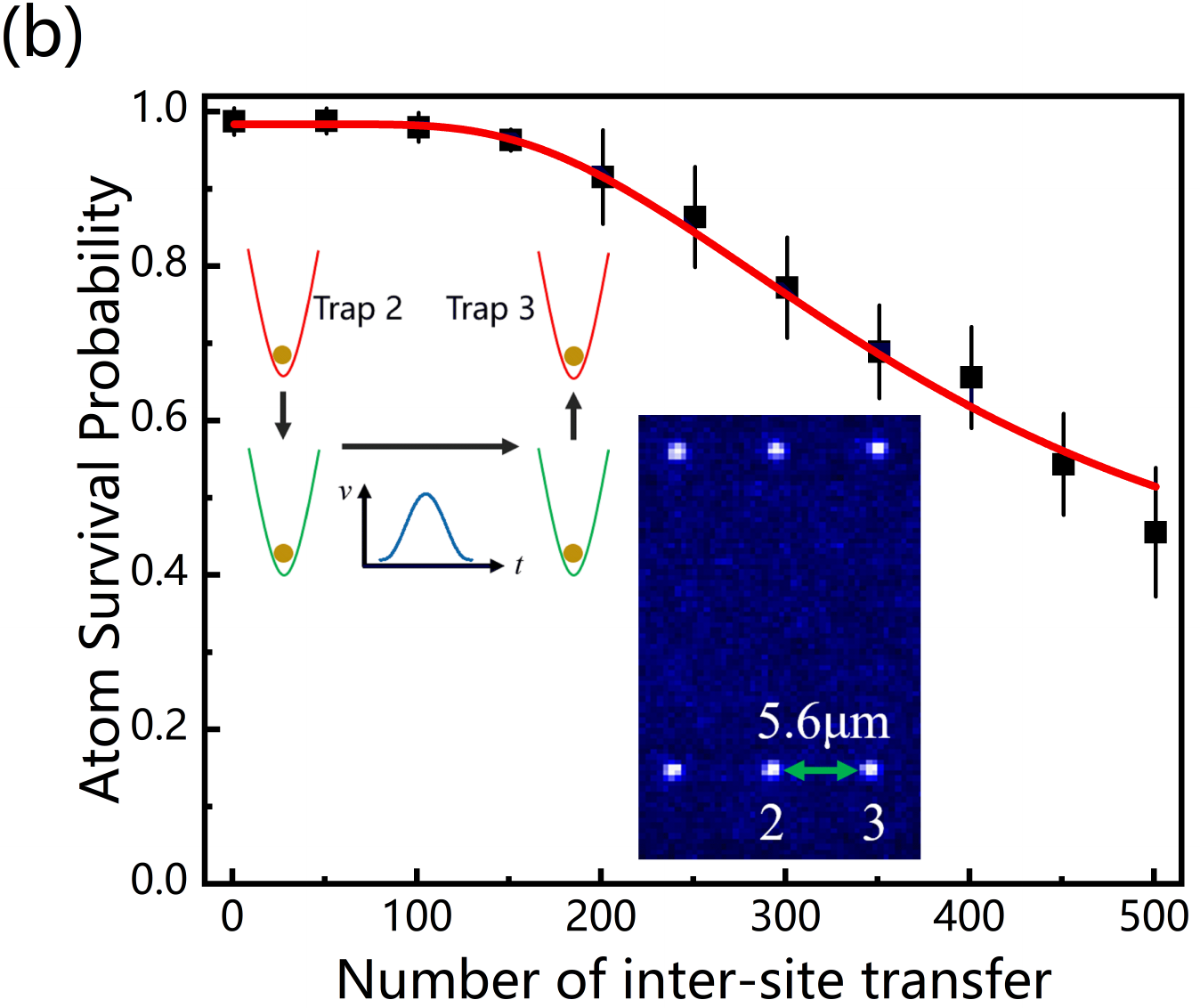}\hfill
    \includegraphics[width=0.41\textwidth]{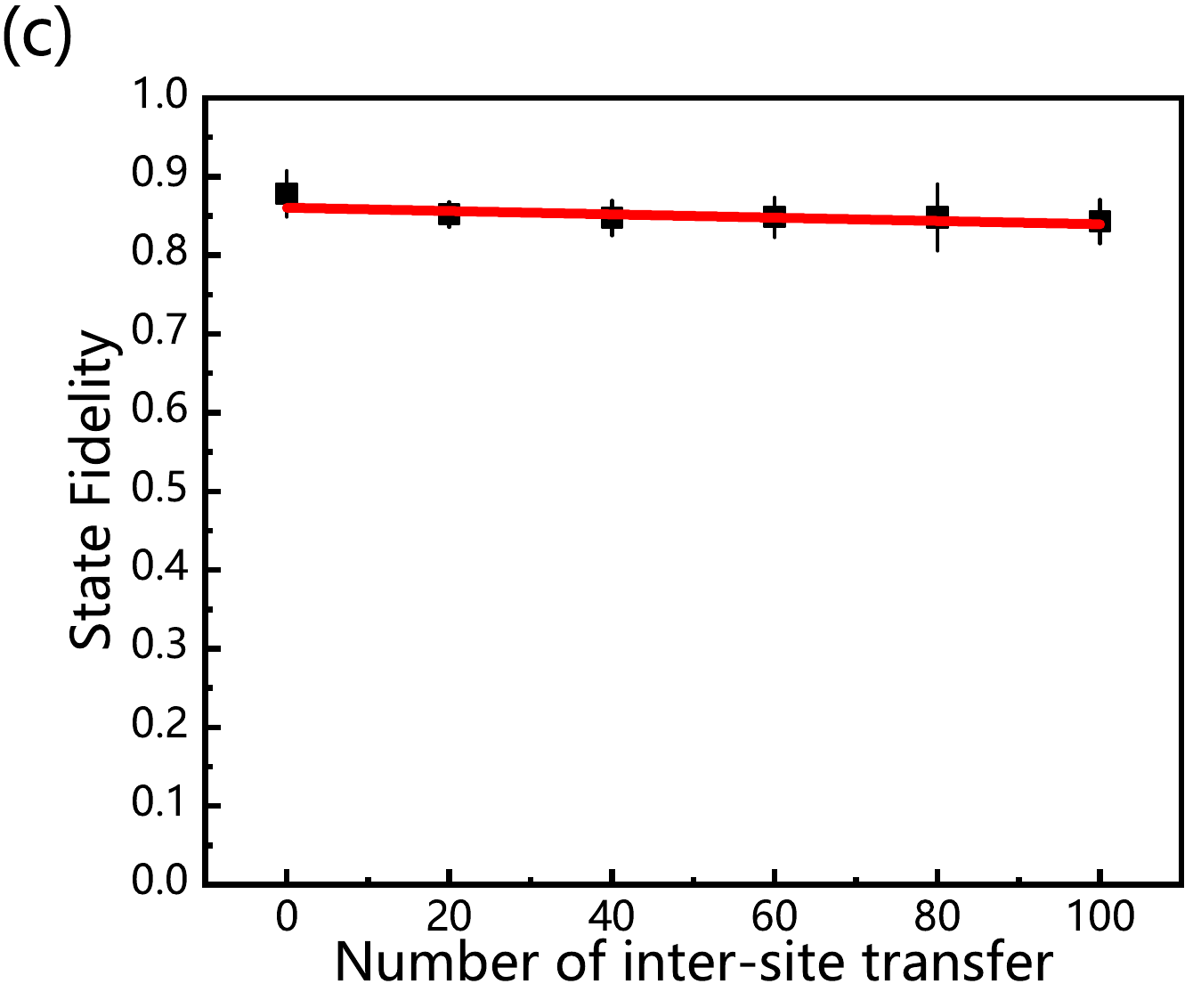}
    \caption{Fast coherent inter-site transfer of a single atomic qubit.
    (a) Measured survival probability with different transport time. The data are fitted using a heating scaling law $\Delta T_{\text{transport}} \propto t^{-p}$. The free fit yielding $p=8.48 \pm 0.79$ (black curve) is consistent with the theoretical $t^{-8}$ prediction (red curve). Inset, schematic of long distance transport cycle.
    (b) Atom survival probability after different numbers of moves between the Trap2 and Trap3. Inset, the detailed qubit move procedure comprises two transfer processes and one transport process.
    (c) Internal-state fidelity after different numbers of moves fitted with exponential decay function.}
    \label{fig:3}
\end{figure}

\section{Fast Coherent inter-site Transfer of Atomic Qubits}

To extend our fast, low-heating in situ transfer to inter-site operations between separate S-traps, an additional spatial transport step is required. We analyze the transport-time dependence using long-distance transport measurements to determine the optimal transport time. An atom is loaded into the M-trap, and the M-trap is then moved back and forth over 35 $\mu \text{m}$ for a total 51 one-way transport segments. As shown in \cref{fig:3}(a), the survival increases with transport time and saturates near 130 $\mu \text{s}$. For our STA trajectory (Eq. 1), the transport-induced heating is expected to scale as $t^{-8}$ derived in the Supplemental Material. To quantify the time dependence, we assume $\Delta T_{\text{transport}}=At^{-p}$ and substitute it into Eq. (3) to fit data, yielding an exponent of $p = 8.48 \pm 0.79$ in the free fit, consistent with the $t^{-8}$ scaling in the Supplemental Material.

To verify the adaptability of the STA transfer waveform across different transport distances, we then demonstrate the full inter-site coherent transfer over a short distance within the array. We selected two adjacent traps (Trap2 and Trap3) separated by $5.6\ \mu\text{m}$ in the tweezer array (see \cref{fig:fig1}), and repeatedly transfer the atom between them.  The moving trajectory of M-trap follows Equation (1) (see inset of \cref{fig:3}(b)), with a transport time of $100\ \mu\text{s}$. The temperature of atom gradually increases with the transfer count, after $500$ inter-site transfers, the atom loss approached 0.5. The single inter-site transfer heating rate is $0.783(17)$ $\mu\text{K}$, as determined by fitting the experimental data to Equation (3), and induces a reduction of $\sim1\%$ in the radial motional ground-state fraction after the first inter-site transfer. Since one complete inter-site transfer consists of two in situ transfer steps and one transport step, subtracting the measured in situ heating of both traps yields a single transport process heating rate of $0.627\ \mu\text{K}$, which induces a reduction of $\sim0.8\%$ in the radial motional ground-state fraction after the first inter-site transfer. As before, we use quantum state tomography to characterize the inter-site transfer fidelity,  the single-transfer fidelity is extracted to be 0.9998(1) (see \cref{fig:3}(b)).

\begin{figure}[t]
    \centering
    \includegraphics[width=0.48\textwidth]{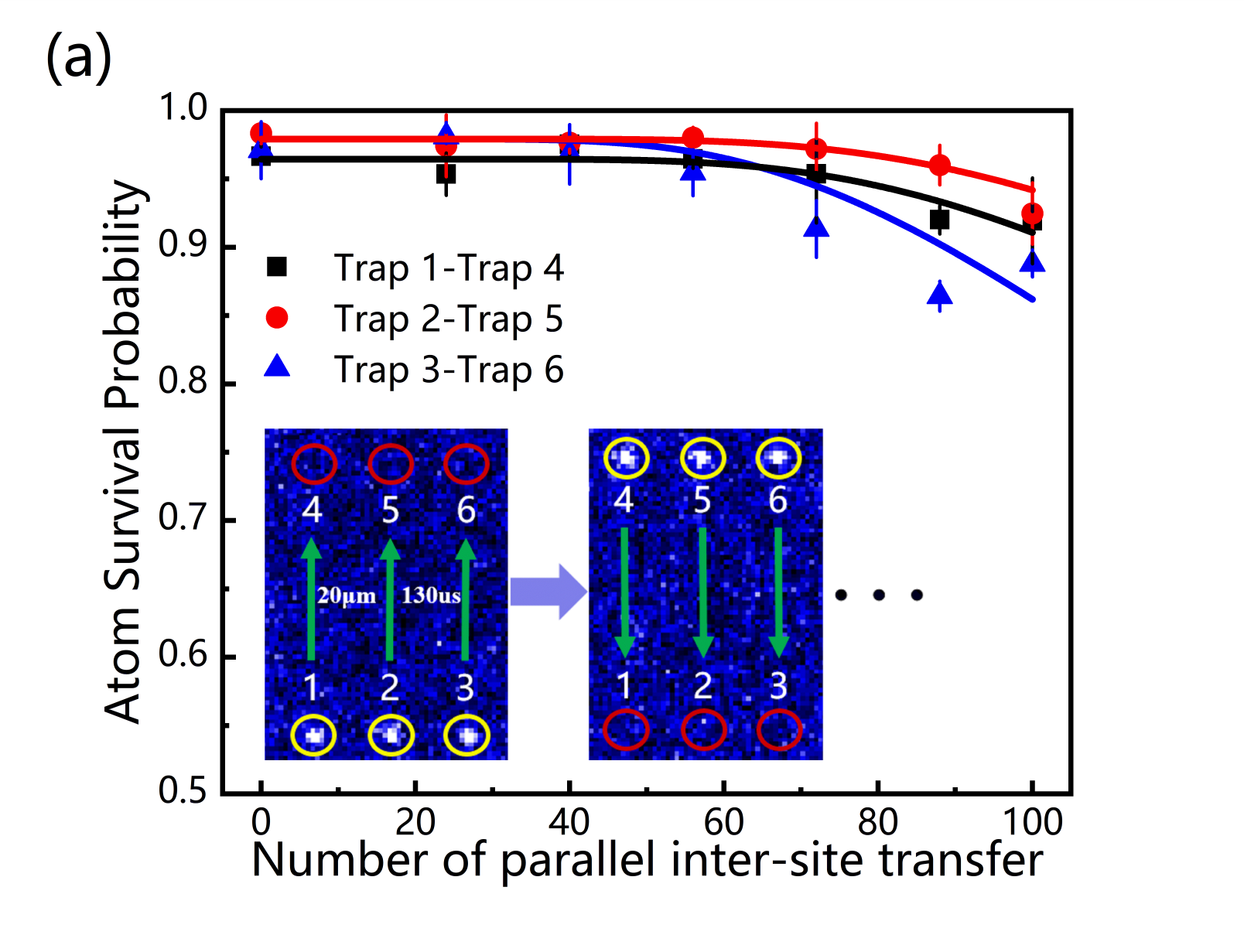}\hfill
    \includegraphics[width=0.48\textwidth]{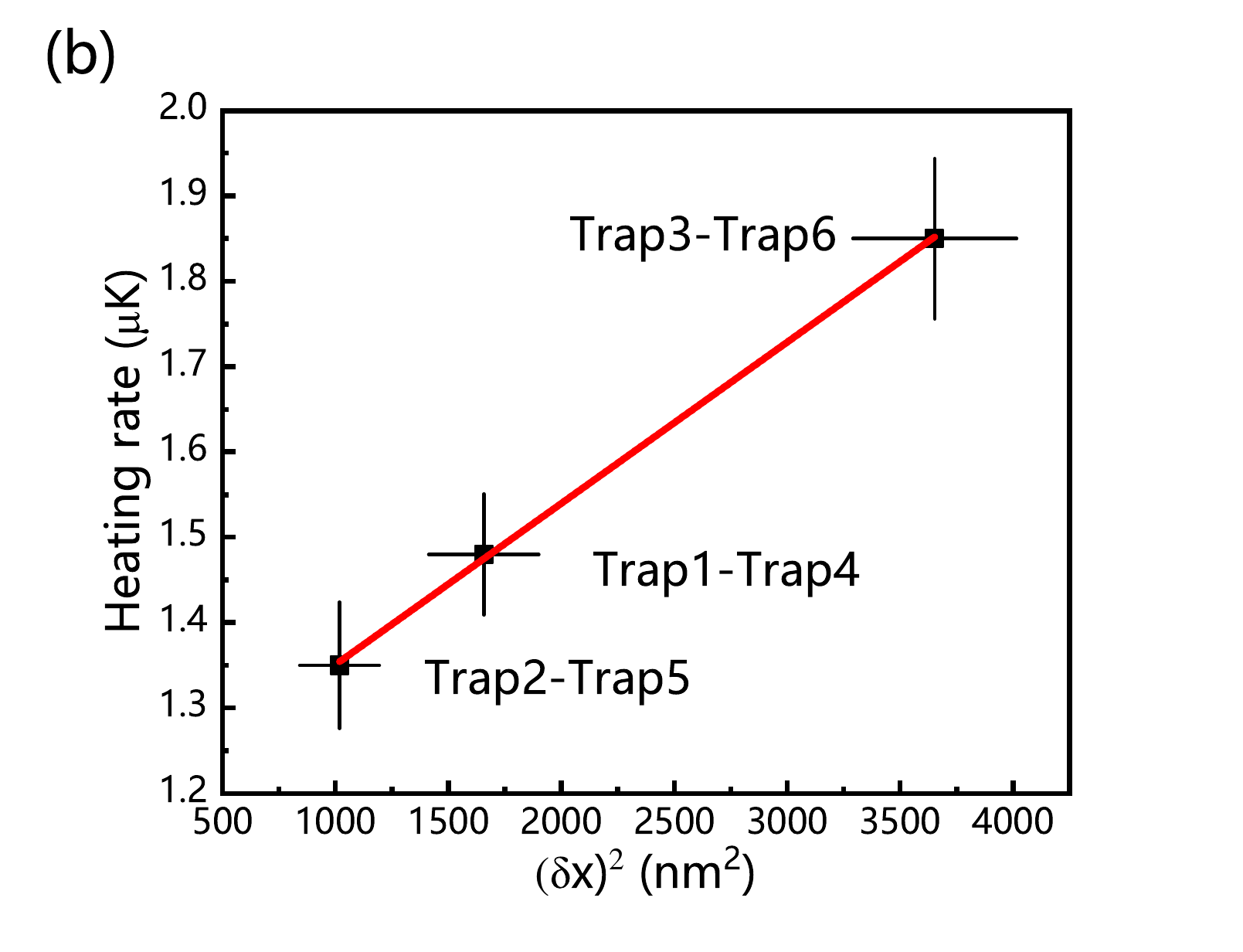}
    \caption{Parallel inter-site transfer of atomic qubits.
    (a) Atom survival probability after different numbers of parallel inter-site transfers.
    (b) The heating rates per inter-site transfer for the three groups, which are caused by different misalignment distances.}
    \label{fig:4}
\end{figure}

The ability to support parallel qubit transfer within the array is essential for quantum processor to enhance circuit efficiency. We use an AOD to generate three M-traps arranged horizontally with an equal spacing of $5.6\ \mu\text{m}$. It can simultaneously match Trap1\textasciitilde Trap3 or Trap4\textasciitilde Trap6(shown in \cref{fig:fig1}). Then, we transfer three qubits from the center to the edge of the fiber array. Due to the size of the fiber array, the moving distance is only $20\ \mu\text{m}$, and the total single inter-site transfer time is $150\ \mu\text{s}$.  The heating rates per inter-site transfer for the three groups are $1.48(7)  ~\mu\text{K}$~(Trap1-Trap4), $1.35(7) ~ \mu\text{K}$~(Trap2-Trap5) and  $1.89(9)~\mu\text{K}$~(Trap3-Trap6), respectively (see \cref{fig:4}(a)), which induce reductions of $\sim1.8\%$, $\sim 1.7\%$ and $\sim 2.3\%$ in the radial motional ground-state fraction after the first inter-site transfer. We attribute this excess heating to a misalignment between the S-trap and M-trap, which arises from fabrication-induced non-uniformity in the fiber array, leading to inconsistent trap spacing.

To understand and optimize the heating rate, we model the complete inter-site transfer physically as three sequential steps: in situ transfer, transport, and in situ transfer. The total heating is expressed as:
\begin{equation}
\begin{aligned}
\Delta T &= \Delta T_{\text{transfer}}^{(1)} + \Delta T_{\text{transport}} + \Delta T_{\text{transfer}}^{(2)} \\
&= T_{\text{basic}}^{(1)} + \Delta T_{\text{mis}}^{(1)} + T_{\text{transport}}(D,t) \\
&\quad + T_{\text{basic}}^{(2)} + \Delta T_{\text{mis}}^{(2)}\notag
\end{aligned}
\end{equation}where $\Delta T_{\text{basic}}^{(1)}=\Delta T_{\text{basic}}^{(2)}$  denotes the basic heating of well aligned in situ transfer step, $\Delta T_{\text{mis}}^{(i)}$ denote the misalignment-induced heating from the in situ transfers, and $\Delta T_{\text{transport}}$ corresponds to the heating generated during the M-trap motion. For the well aligned case, we take $\Delta T_{\text{basic}}^{(1)}+\Delta T_{\text{basic}}^{(2)}=0.156(9)\mu \text{K}$, which is the average of the two previously measured in situ heating rates. Under the harmonic approximation, a spatial misalignment between the S-trap and M-trap introduces a position shift that yields:

\begin{equation}
\begin{aligned}
\Delta T_{\text{mis}}&=\Delta T_{\text{mis}}^{(1)}+\Delta T_{\text{mis}}^{(2)}\\&=\frac{1}{2}\alpha m\omega_{0}^{2}\left[(\delta x_{\text{start}})^{2}+(\delta x_{\text{target}})^{2}\right]\notag
\end{aligned}
\end{equation}where $\alpha$ quantifies the non-adiabaticity of the transfer process and is a parameter dependent on the in situ transfer time. We performed a linear fit of the measured total heating against $(\delta x)^2$ (see Fig. 4(b)), obtaining $\Delta T_{\text{transport}}^{\text{exp}}\approx 1.01~\mu \text{K}$. On the other hand, based on our STA trajectory, the theoretical transport-induced heating follows the scaling $\Delta T_{\text{transport}}\propto D^{2} /(\omega_{0}^7t^8)$. Using the transport heating $\Delta T_{\text{transport}}=0.627 \mu \text{K}$ extracted from the Trap2-Trap3 experiments $({D=5.6 \mu \text{m},t=100\mu \text{s}})$ as a reference, the scaling predicts $\Delta T_{\text{transport}}^{\text{ideal}}\approx 0.98\mu \text{K}$ for the parallel configuration $({D=20\mu \text{m},t=130\mu \text{s}})$, consistent with our experimental result, confirming that our model captures the dominant heating mechanisms. Compared to the constant-jerk trajectory with $\Delta T_{\text{transport}}\propto D^{2}/(\omega_{0}^3t^4)$ and the adiabatic sinusoidal trajectory with $\Delta T_{\text{transport}}\propto D^{2} /(\omega_{0}^5t^6)$ employed in several other experimentally realized schemes\ucite{Bluvstein2022,Manetsch2025}, our STA trajectory offers an improved scaling of $\Delta T_{\text{transport}} \propto D^{2} /(\omega_{0}^7t^8)$, which facilitates a further reduction in heating rate for a given transport distance and time.

\section{Summary and Discussion}
In summary, we experimentally demonstrate fast, coherent transfer of atomic qubits in a two-dimensional optical trap array, establishing a key capability for accelerating non-local qubit connectivity. Enabled by independent, site-resolved control of trap depths in a fiber array architecture\ucite{Li2025}, we realize in situ transfer between traps with a success probability exceeding 0.9999 in 10 $\mu\text{s}$, representing faster and higher fidelity transfer compared to previous deep-trap-based adiabatic extraction from a shallow trap \ucite{Yang2016,Bluvstein2024,Reichardt2025,Manetsch2025}. Furthermore, we have experimentally verified a transport trajectory that achieves a lower heating rate and a faster operation speed.

Currently, the heating of the in situ transfer is primarily caused by power fluctuations and interference between the S-trap and M-trap, which can be improved through better power stabilization and larger detuning of the frequencies of the two beams. Moreover, the success probability of parallel inter-site transfer remains lower than single-atom inter-site transfer, primarily due to heating induced by the limited positioning accuracy of individual fibers. And the residual motional excitation after transport can further reduce the fidelity of the following Rydberg-mediated two-qubit gates. To solve this non-uniform trap spacing problem and scale up the atomic array, our next step is to employ femtosecond laser-written 3D photonic waveguides, which have the potential to support more than one thousand control channels.\ucite{Gattass2008,DongQiMa2025}. This approach can reduce spatial non-uniformity to below $10~\text{nm}$ in our experimental setup, thereby enabling parallel inter-site transfer over hundreds of cycles. To further push the inter-site transfer capability to thousands of cycles, the inherent heating of system can be systematically mitigated by enhanced sideband cooling and employing lower-noise, far-detuned traps, quantum optimal  control\ucite{Manolol2021,Torrontegui2011,Bowler2012,Walther2012} and machine-learning-based methods\ucite{Manetsch2025}, ultimately pushing the transfer capability toward the fundamental limits set by background-gas collisions. Furthermore, residual heating is expected to be mitigated in deeper circuits by mid-circuit recooling \ucite{Lin2026} or atom replacement in a fault-tolerant architecture\ucite{Bluvstein2026}. With these capabilities, together with fast, high-fidelity, individually addressed gates\ucite{Li2025,Radnaev2025,BichenZhang2024}, our approach paves the way for a neutral-atom quantum computing architecture that simultaneously supports high quantum error correction cycle rates and efficient entanglement generation between logical qubits.


\section*{Acknowledgements}
This work was supported by the National Key Research and Development Program of China under Grant No. 2021YFA1402001, the	National Innovation Program for Quantum Science and Technology of China under Grant No. 2023ZD0300401, the National Natural Science Foundation of China under Grants No. 12004397, No. 12261131507, No.12074391, No.U22A20257, No. 12121004 and No. 12241410,the CAS Project for Young Scientists in Basic Research under Grant No. YSBR-055, the Major Program (JD) of Hubei Province under Grant No. 2023BAA020.

\clearpage
\appendix
\renewcommand{\thefigure}{S\arabic{figure}}  
\setcounter{figure}{0}
\setcounter{table}{0}
\renewcommand{\thetable}{S\arabic{table}}
\section*{Supplemental material}
\section{Theoretical Derivation of Atomic Heating for Different Transport Trajectories}

To understand the influence of different transport trajectories on atomic heating and to justify the selection of the Bernstein polynomial trajectory in this work, we provide a detailed theoretical derivation. Based on the harmonic oscillator approximation and the adiabatic approximation, when an atom moves along a designed trajectory $x(t)$ in an optical trap, the average increase in vibrational quantum number $\Delta N$ is determined by the amplitude of the Fourier spectrum of the motion's acceleration at the trap frequency $\omega_0$ \cite{Carruthers1965,Manetsch2025}:

\begin{equation}
\Delta N = \frac{|\tilde{a}(\omega_0)|^2}{(2 x_{\text{zpf}} \omega_0)^2}, \quad \text{where} \quad x_{\text{zpf}} = \sqrt{\frac{\hbar}{2m\omega_0}}.
\end{equation}

Here, $\tilde{a}(\omega)$ is the Fourier transform of the acceleration $a(t) = d^2x/dt^2$, and $m$ is the atomic mass. Under the experimental condition $\omega_0 t_{\text{total}} \gg 1$, the term $|\tilde{a}(\omega_0)|^2$ in Eq.~(S1) is dominated by the leading term in the high-frequency asymptotic expansion of $\tilde{a}(\omega)$ as $\omega \to \infty$. This allows us to compare the heating scaling laws of different schemes by analyzing the mathematical form of the trajectory functions.

To compare the heating rates of different trajectories, we perform an asymptotic analysis on the Fourier transform of the acceleration $\tilde{a}(\omega)$ in the limit $\omega \to \infty$. Under the condition $\omega_0 t_{\text{total}} \gg 1$, the value of $|\tilde{a}(\omega_0)|^2$ in Eq.~(S1) is determined by the term with the lowest power of $\omega$ in the asymptotic expansion.

For the constant-jerk trajectory, defined by the displacement function
$x_{1}(t) = D[3(t/ t_{\text{total}})^2 - 2(t/ t_{\text{total}})^3], ~~t\in[0,  t_{\text{total}}]$,
the leading asymptotic term is proportional to $D/( t_{\text{total}}^2\omega)$. Substituting into Eq.~(S1) yields a scaling of $\Delta N \propto D^2/(\omega_0^3  t_{\text{total}}^4)$.

For the adiabatic sinusoidal trajectory
$x_{2}(t) = (D/2)[(1/\pi)\sin(\pi \tau) + \tau + 1], ~~\tau=2t/ t_{\text{total}}-1 \in [-1,1]$,
the leading term scales as $D/( t_{\text{total}}^3\omega^2)$, corresponding to $\Delta N \propto D^2/(\omega_0^5  t_{\text{total}}^6)$.\cite{Manetsch2025}

For the optimized Bernstein polynomial trajectory used in this work,
$x_{3}(t) = D[-20(t/ t_{\text{total}})^7 + 70(t/ t_{\text{total}})^6 -84(t/ t_{\text{total}})^5 + 35(t/ t_{\text{total}})^4]$,
the leading asymptotic term scales as $D/( t_{\text{total}}^4\omega^3)$, leading to a heating scaling of $\Delta N \propto D^2/(\omega_0^7  t_{\text{total}}^8)$. This demonstrates that by designing smoother trajectories, the motional heating can be suppressed by an additional factor of $1/(\omega_0  t_{\text{total}})$.

\begin{figure}[t]
    \centering
    \begin{minipage}[b]{0.9\linewidth}
        \centering
        \includegraphics[width=\linewidth]{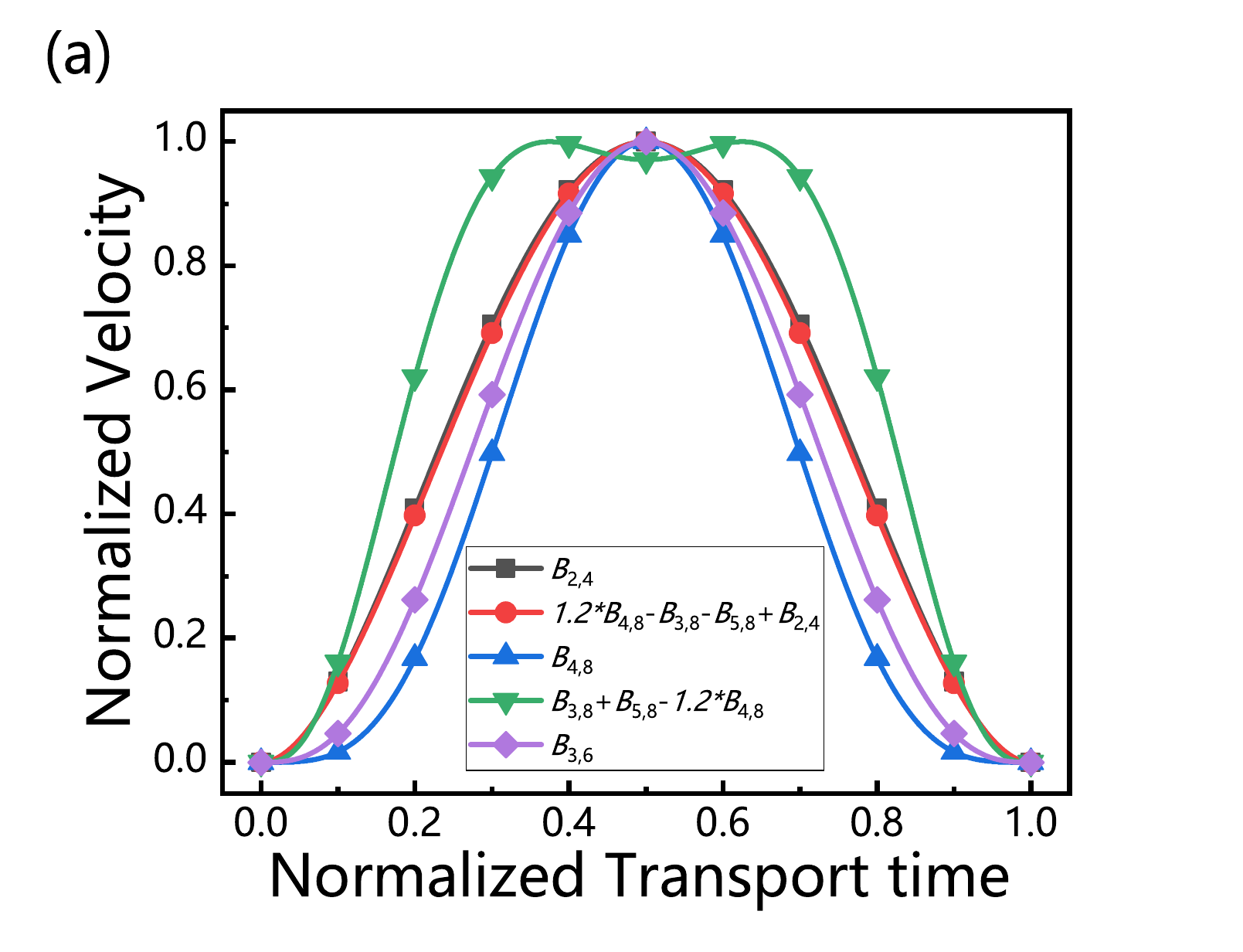}
    \end{minipage}
    \hfill
    \begin{minipage}[b]{0.9\linewidth}
        \centering
        \includegraphics[width=\linewidth]{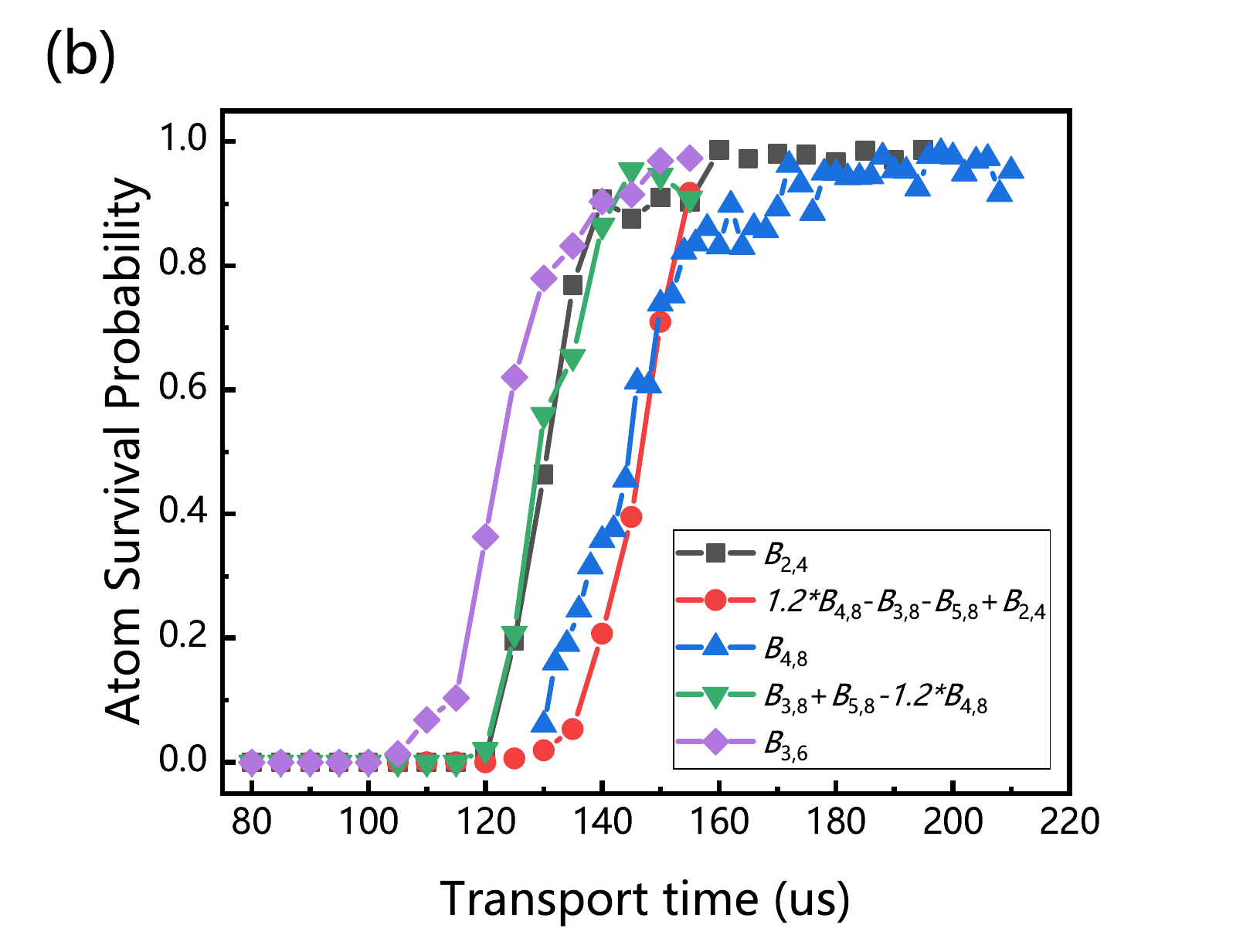}
    \end{minipage}
    \caption{(a) The different normalized velocity-time relationship. (b) The relationship between atomic survival probability and transport time with different trajectories.}
    \label{fig1}
\end{figure}

Our work theoretically demonstrates the significant advantage of the Bernstein polynomial trajectory in suppressing motional heating. To validate the theory and determine optimal parameters, we experimentally compared the atom survival probabilities for various Bernstein polynomial trajectories of different orders during long-distance transport. As shown in \cref{fig1}, an atom is  loaded into the M-trap, and the M-trap is then moved back and forth over 35 $\mu \text{m}$ for a total 51 one-way transport segments. The atom survival probability varies significantly with transport time for different trajectories. The experimental results show that the trajectory derived from the coefficients $B_{3,6}$ (i.e., the function $x_{3}(t)$ above) achieves the highest atom survival probability for a given transport time $T$.

\section{Effective transport induced by amplitude exchange under residual misalignment}

In the in situ transfer, the S-trap and M-trap ideally overlap spatially, and the atom is transferred by exchanging their trap depths. Let the S-trap center be at $x=0$ and the M-trap center be at $x=\delta x$. In the small-misalignment limit and near the trap minimum, the combined potential can be approximated as
\begin{equation}
V(x,t) \simeq \frac{1}{2}m\omega_S^2(t)x^2+\frac{1}{2}m\omega_M^2(t)\bigl(x-\delta x\bigr)^2+\mathrm{constant}.
\end{equation}
The minimum of the combined potential is therefore located at
\begin{equation}
x_c(t)=\frac{\omega_M^2(t)}{\omega_S^2(t)+\omega_M^2(t)}\,\delta x.
\end{equation}

For matched beam waists and nearly identical trap curvatures, one has $\omega_i^2(t)\propto U_i(t)$, where $U_i(t)$ is the trap depth. If the total depth is kept constant during the transfer,
\begin{equation}
U_S(t)+U_M(t)=U_0,
\end{equation}
and we write
\begin{equation}
U_M(t)=U_0\lambda(t),\qquad U_S(t)=U_0[1-\lambda(t)],
\end{equation}
then the effective trap minimum becomes
\begin{equation}
x_c(t)=\lambda(t)\,\delta x.
\end{equation}
Therefore, a residual spatial misalignment converts the amplitude-exchange process into an effective short-distance transport of the trap minimum. In the ideal case where $\delta x=0$, one has $x_c(t)=0$, and the depth exchange itself does not induce motional excitation.

The corresponding effective acceleration is
\begin{equation}
a_c(t)=\ddot{x}_c(t)=\delta x\,\ddot{\lambda}(t).
\end{equation}
Substituting this effective trajectory into Eq.~(A1), the heating induced by the residual misalignment is determined by the Fourier component of $a_c(t)$ at the trap frequency. Therefore, once the modulation function $\lambda(t)$ (which corresponds to $x(t)$ defined in Eq.~(1) of the main text) is chosen, the transfer heating can be analyzed as a transport problem over the distance $\delta x$. In particular, if $\lambda(t)$ is chosen to have the same smooth boundary conditions as an STA transport trajectory, the corresponding misalignment-induced heating follows the same scaling law as that transport trajectory, with the transport distance $D$ replaced by $\delta x$.

\section{Experiment Setup and Measurement of Static Trap Misalignment in Parallel Transfer}

The experimental setup of the fiber-array-based atomic quantum computer has been described in detail in our previous work.\cite{Li2025} Each fiber serves as the interface for trapping and addressing an individual $^{87}\text{Rb}$ atom, simultaneously enabling scalability and independent control. In this work, we utilize fiber-array-based optical traps as S-Traps, each with independently controllable depth, and introduce an M-Trap for qubit transport within the array. The M-Trap laser shares the same source as the dipole light for the fiber array, with a wavelength of 830 nm. 

\begin{figure}[!h]
    \centering
    \includegraphics[width=0.36\textwidth]{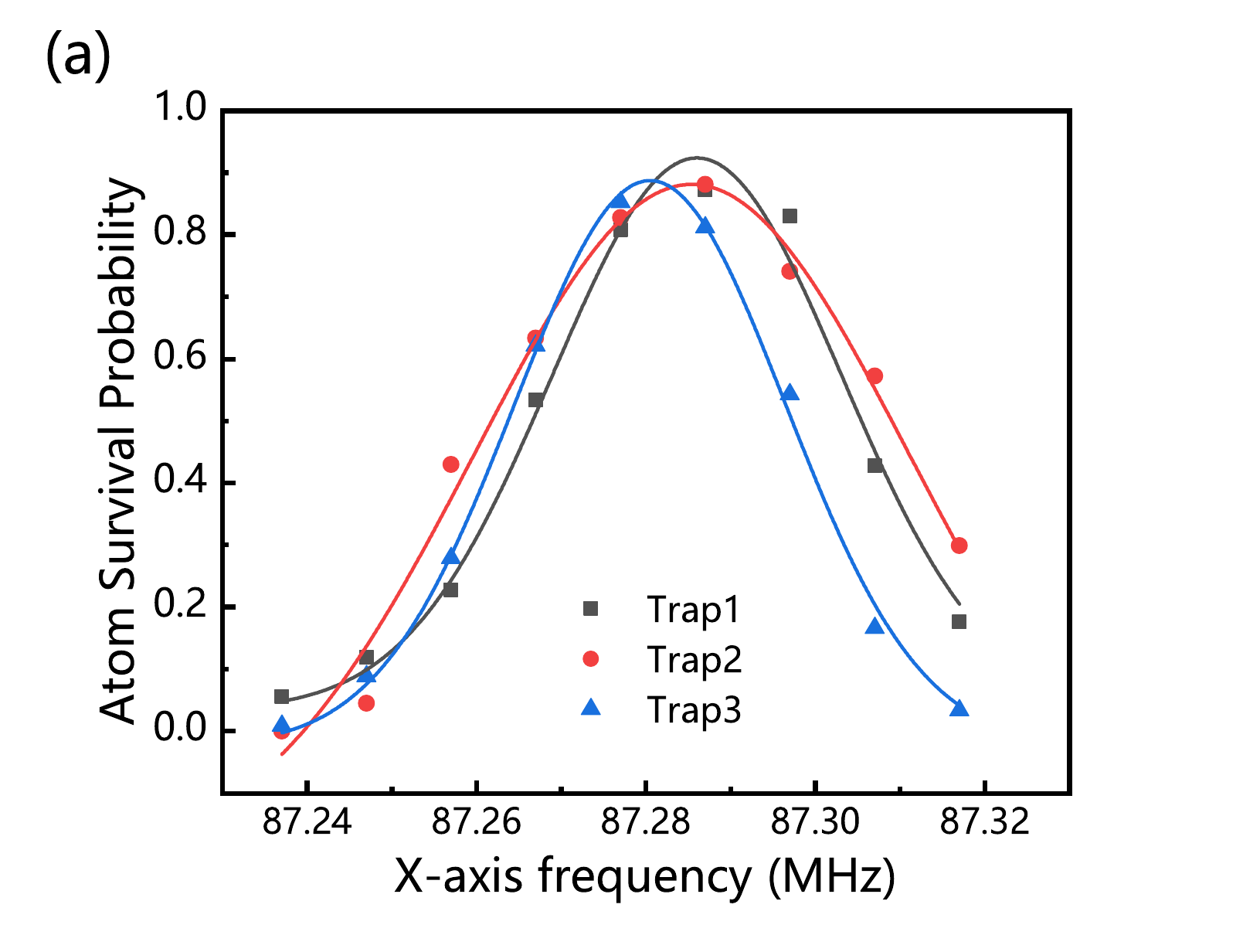}
    \includegraphics[width=0.36\textwidth]{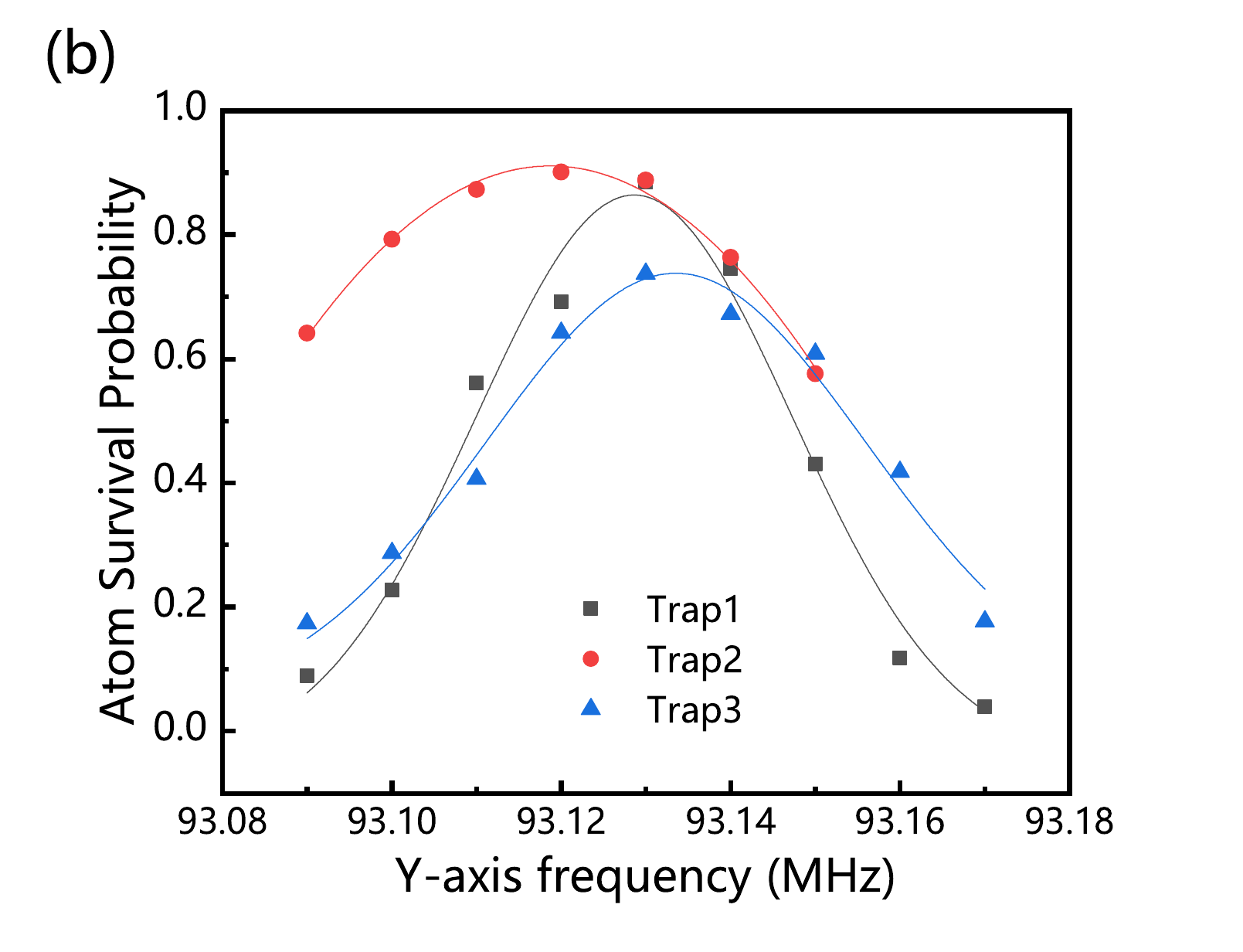}
    \includegraphics[width=0.36\textwidth]{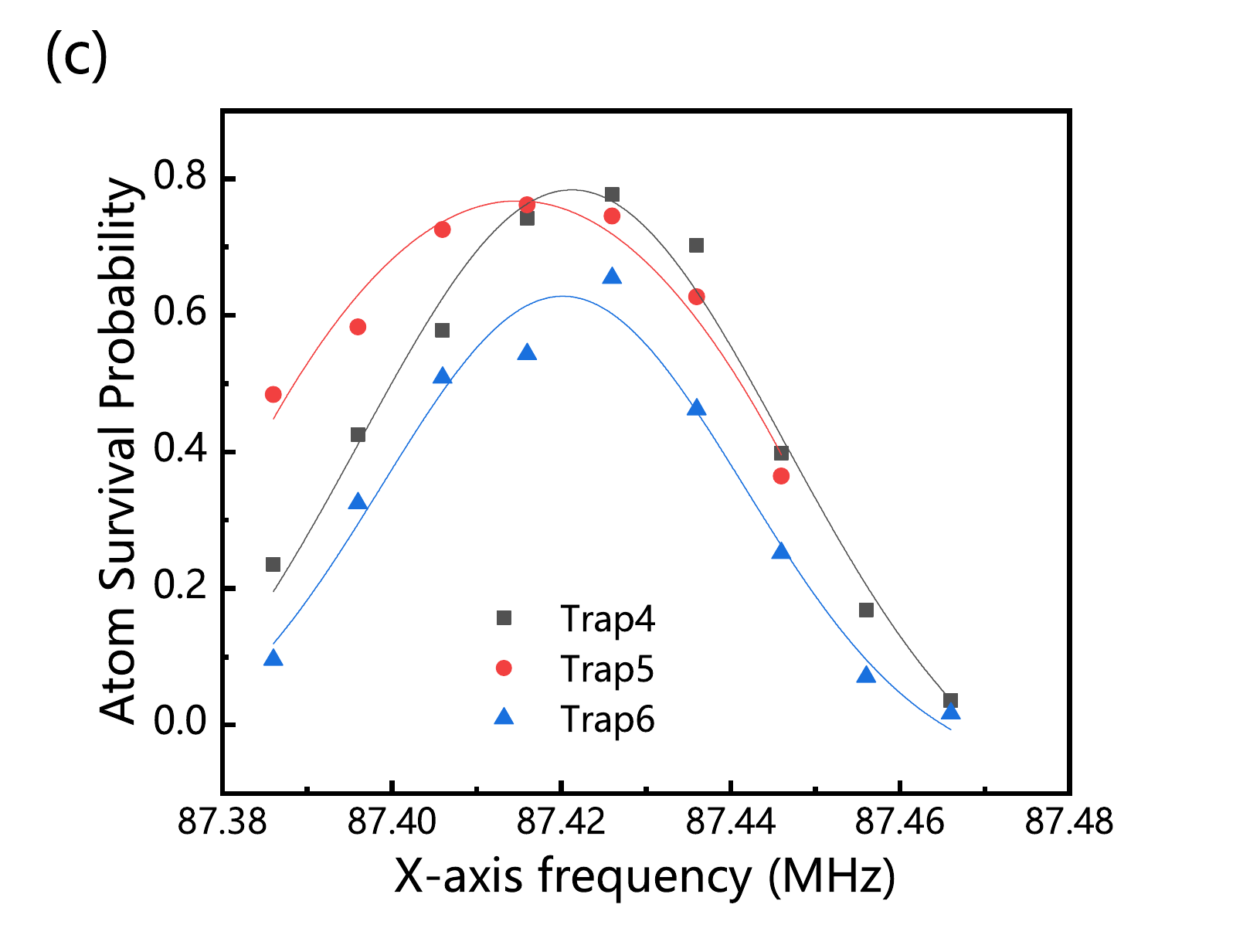}
    \includegraphics[width=0.36\textwidth]{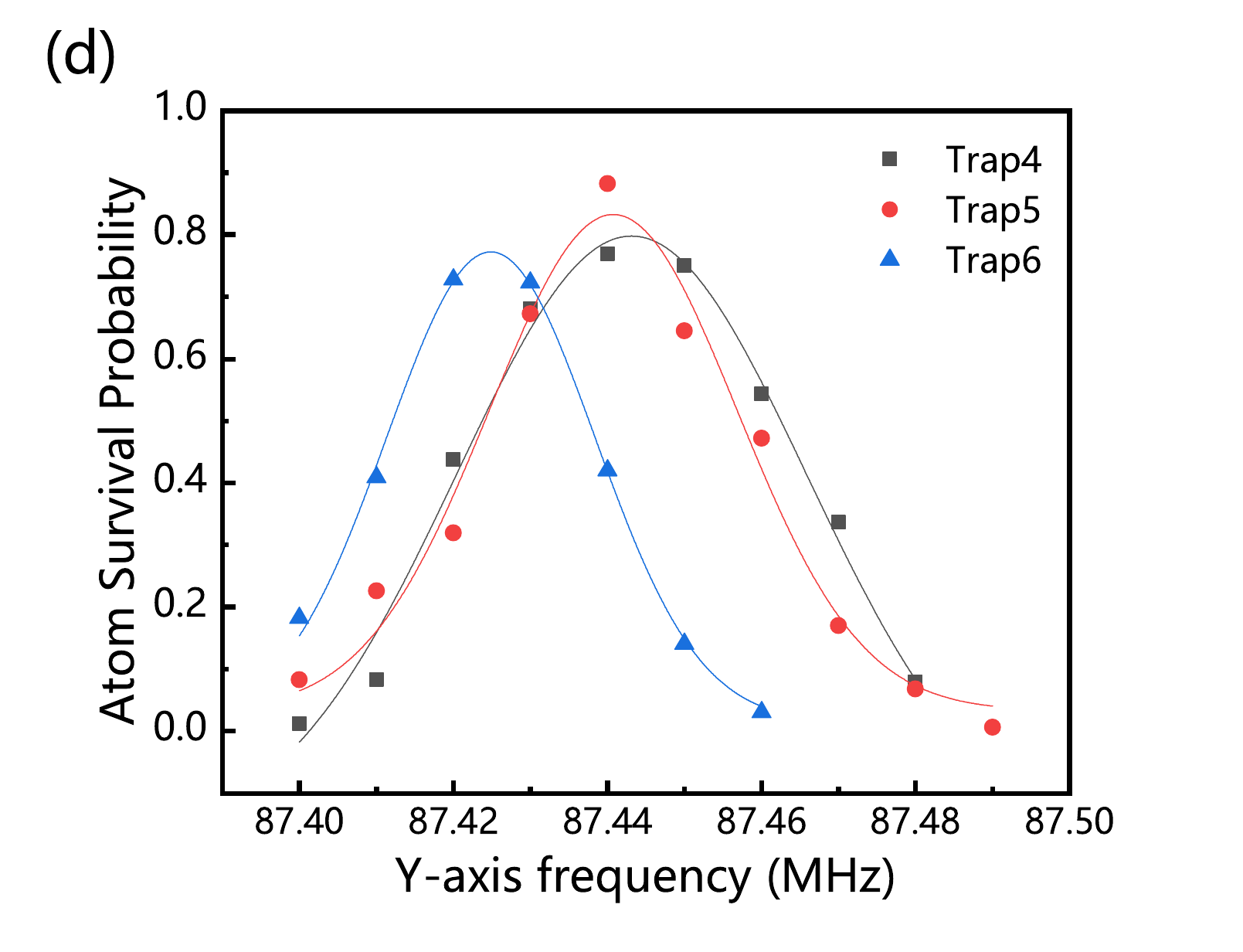}
    \caption{Measurement of optical center positions for the six static traps. (a) X-direction scan results for Traps 1--3. (b) Y-direction scan results for Traps 1--3. (c) X-direction scan results for Traps 4--6. (d) Y-direction scan results for Traps 4--6. Each data point represents the atom survival probability at a specific AOD frequency, and the 2D Gaussian fit yields the optimal transfer frequency corresponding to the optical center of each trap.}
    \label{Fig:s2}
\end{figure}

The intensity of M-Trap  is modulated by an acousto-optic modulator (AOM), using a amplitude modulated radio-frequency (RF) signal generated from an arbitrary waveform generator (AWG) (CIQTEK AWG4100). The position of  M-trap is modulated by a two-dimensional acousto-optic deflector (AOD), the frequency-tuning signal for the AOD is generated by another AWG  (Keysight M3202A). After emerging from the AOD, the M-Trap beam first passes through a reflective 4f relay system with $1:1$ magnification, and then combined with the S-Trap beam using a beam splitter ($T:R = 9:1$). To improve the transfer success rate, the combined M-Trap and S-Trap beams are configured to have the same beam waist. The combined beams subsequently pass through a linear polarizer (LP)  and a a liquid crystal variable retarder (LCVR), which are employed to refine and control their polarizations dynamically. Finally, after the combined beams pass through another 4f beam-expanding system ($\times$5), the beam is focused by an objective into the vacuum chamber, where it forms optical tweezers.

Fast, low-heating in situ coherent transfer of atomic qubit requires precise alignment of the M-trap and S-trap in both the radial and axial directions which needs iterative optimization. For the radial direction, the M-trap position is scanned by varying the AOD's RF frequency. First, we coarsely align the M-trap and S-trap based on the fluorescence images of atoms on the sCMOS camera.Then, we precisely scan the position of the M-trap around each S-trap. During the scan, the atom was transferred for 100 cycles between the S-trap and M-trap. The coordinates of the S-trap are obtained by fitting the atom survival probability as a function of position. In our system, the central radio frequency of the AOD is 90 MHz, and a change of 0.288 MHz corresponds to a shift of the trap position by 1 $\mu$m.  For the axial direction, we finely scan the M-Trap position by translating a mirror within the reflective 4f relay system along the optical axis. Finally, we measure the resonance frequency of the M-trap and  S-trap, the difference was found to be less than 3\%. By the way, to mitigate drift in the M-trap position caused by temperature fluctuations in the AOD crystal, we multiplex another sinusoidal RF signal with the AWG signal to maintain a constant crystal temperature during idle periods.

To understand the heating during atomic transfer and to validate the position-mismatch model established in Section 4.2 of the main text, it is necessary to measure the alignment error between the static traps $S_i$ created by the fiber array and the moving traps $M_i$ created by the acousto-optic deflector.

\cref{Fig:s2} presents the calibration results for all six static traps. Each static trap $S_i$ ($i=1 \sim 6$) has its spatial position determined by a fixed RF frequency pair $(f_x, f_y)$ driving the AOD. Due to fabrication tolerances of the fiber array and optical aberrations, their actual central positions $(X^S_i, Y^S_i)$ require precise calibration. The calibration method is as follows: atoms are loaded into the target static trap $S_i$, and an in-situ transfer operation to itself is performed. During this process, the position of the dynamic trap $M_i$, determined by the AOD's RF frequency pair $(f_x, f_y)$, is scanned over a two-dimensional grid within a $\sim$ 200 nm  range. For each $M_i$ position, we perform 100 transfer cycles and measure the final atom survival probability. When the $M_i$ position aligns with the center of $S_i$, the additional heating due to position mismatch is minimized, resulting in the highest average atom survival probability after multiple transfers. The frequency pair $(f_x, f_y)$ and the corresponding survival probability for each scan point are recorded. The survival data is then fitted with a 2D Gaussian function in frequency space. The center of the fitted Gaussian distribution, $(f_x^S, f_y^S)$, yields the optimal transfer frequency for that static trap $S_i$, corresponding to its optical center in AOD frequency coordinates. In our system, frequency differences are converted to spatial displacements $\delta x_i$ using a conversion factor of $1.615\ \text{MHz} \leftrightarrow 5.6 \mu m$.
\begin{table*}[t]
\centering
\small
\setlength{\tabcolsep}{14pt}
\begin{tabular}{cccc}
\hline \hline 
Trap Label & $X^S$ (MHz)/$Y^S$ (MHz) & $X^M$ (MHz)/$Y^M$ (MHz) & Displacement Mismatch $\delta x$ (nm) \\
\hline 
$S_1$ / $M_1$ & 87.2860(7) / 93.1286(9) & 87.286 / 93.123 & 19(3) \\
$S_2$ / $M_2$ & 88.8984(9) / 93.1187(5) & 88.896 / 93.123 & 17(2) \\
$S_3$ / $M_3$ & 90.5065(4) / 93.134(1)  & 90.506 / 93.123 & 38(3) \\
$S_4$ / $M_4$ & 87.4212(9) / 87.4431(8) & 87.419 / 87.433 & 36(3) \\
$S_5$ / $M_5$ & 89.030(1) / 87.4407(9)  & 89.029 / 87.433 & 27(3) \\
$S_6$ / $M_6$ & 90.650(1) / 87.4249(4)  & 90.639 / 87.433 & 47(3) \\
\hline \hline 
\end{tabular}
\caption{Measurement of alignment errors between static and transfer traps. The numbers in parentheses reflect uncertainties from the Gaussian fits. The measured mismatch values (17--47~nm) are consistent with the theoretical error range (approx. 50~nm) arising from fiber array fabrication and optical system aberrations.}
\label{tab:alignment}
\end{table*}

    In a single transport, the squared mismatch displacement $\delta^2$ experienced by the atom is the sum of the squares of the mismatches of the start and target traps: $\delta^2 = (\delta x_{\text{start}})^2 + (\delta x_{\text{target}})^2$. This yields:
\begin{align*}
\text{Trap 1}\to\text{Trap 4}&: \delta^2 = (19\ \text{nm})^2 + (36\ \text{nm})^2 = 1657(244)\ \text{nm}^2, \\
\text{Trap 2}\to\text{Trap 5}&: \delta^2 = (17\ \text{nm})^2 + (27\ \text{nm})^2 = 1018(176)\ \text{nm}^2, \\
\text{Trap 3}\to\text{Trap 6}&: \delta^2 = (38\ \text{nm})^2 + (47\ \text{nm})^2 = 3653(363)\ \text{nm}^2.
\end{align*}

    In the parallel transport experiments, we measured the heating rate $\Delta T$ for different trap pairs due to transport cycles with static position mismatch. The heating rate is proportional to $(\delta x)^2 $.
\begin{figure}
    \centering
    \includegraphics[width=0.5 \textwidth]{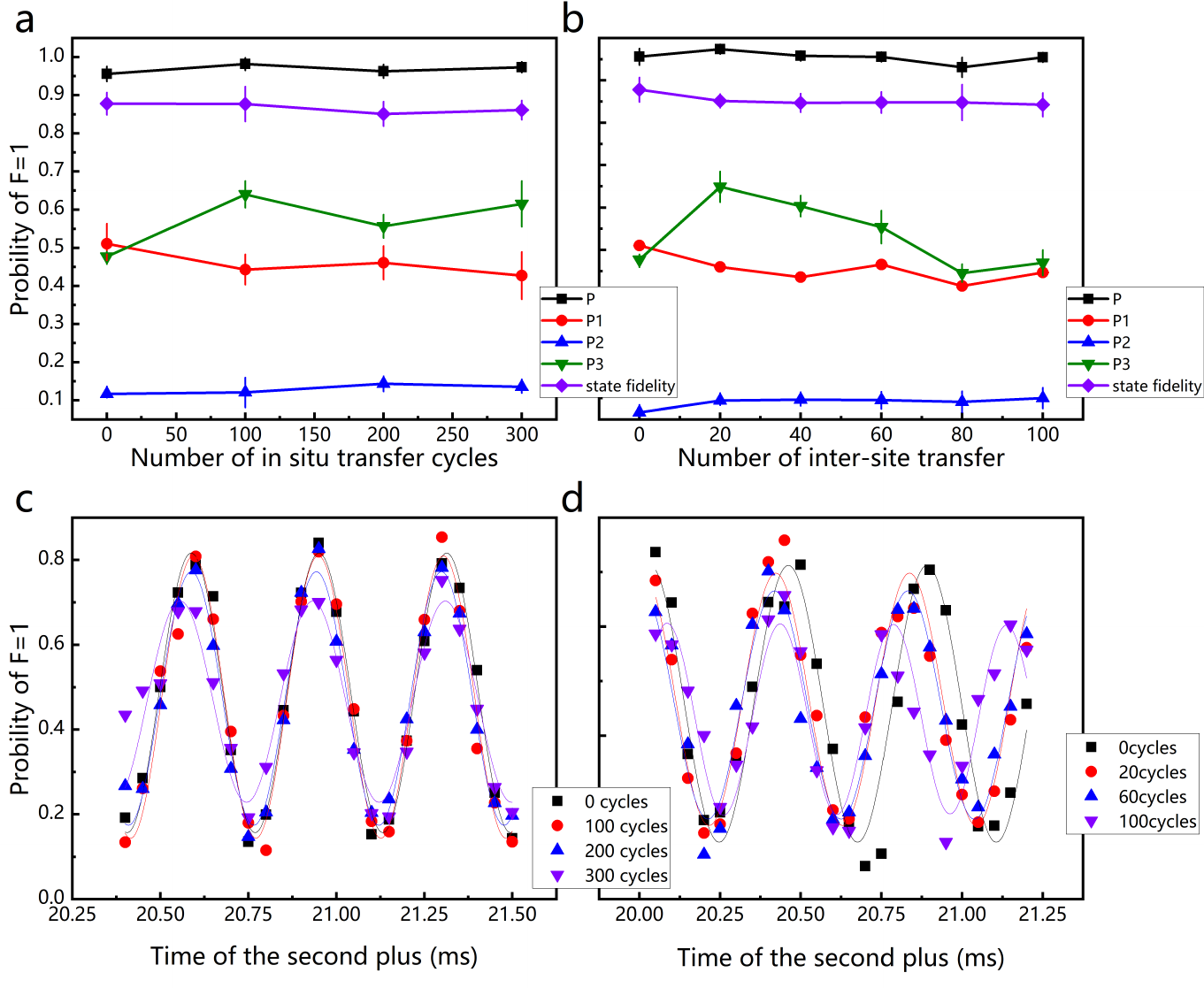}
    \caption{High-fidelity coherent transfer of a single atomic qubit. (a) Atomic qubit state fidelity after different numbers of transfers. (b) Atomic qubit state fidelity after different numbers of transports. (c) Ramsey fringes after different numbers of transfers. (d) Ramsey fringes after different numbers of transports.}
    \label{fig:s3}
\end{figure}
\section{Characterization of Single-Atom Qubit Internal State Coherence}

To characterize the fidelity and coherence preservation of the internal state during the transport process, we employed quantum state tomography and measured Ramsey interference fringes.\cite{Yu2014} This section details the experimental methods and results, with specific data shown in \cref{fig:s3}.

We encode a qubit in the hyperfine ground states $|F=1, m_F=0\rangle$ and $|F=2, m_F=0\rangle$ of $^{87}\text{Rb}$. To extend qubit coherence time, we implement magic-intensity optical trapping technology\cite{Yang2016} and spin-echo sequence.

The specific sequence of state tomography is as follows: the initial state is prepared to $|1,0\rangle$; a $\pi/2$ microwave pulse prepares the atom along the $Y$-axis of the Bloch sphere; after a 10 ms delay, a $\pi$ pulse is applied to suppress inhomogeneous broadening; following another 10 ms delay, a second $\pi/2$ analysis pulse is applied. The phase of this pulse is varied (0, $\pi/2$, $\pi$) to measure the projections $P_1, P_2, P_3$ onto three orthogonal axes of the Bloch sphere. The quantum state is finally read out via state-dependent fluorescence imaging.

To visualize the coherence preservation, we measured Ramsey fringes. The sequence is: after initial state preparation, apply a $\pi/2$ pulse; after a free evolution time $t$, interspersed with transfer/transport operations, apply a second $\pi/2$ pulse with a tunable phase; by scanning the phase or time of the second pulse, the population in the $|2,0\rangle$ state is recorded, yielding oscillatory fringes. The fringe contrast reflects the preserved coherence.

\cref{fig:s3}(a) presents the internal state fidelity of the atomic qubit after undergoing a different number of transfer operations. At a fast $10~\mu \text{s}$ swap time, the state fidelity remains at 0.85 even after 300 transfers, indicating nearly lossless internal quantum state.

\cref{fig:s3}(b) shows the internal state fidelity after the atom undergoes different numbers of complete transport operations between two static traps. After 100 transports, the fidelity remains at 0.84.

To more intuitively demonstrate the preservation of quantum coherence, we measured Ramsey interference fringes after inserting different numbers of operations.

\cref{fig:s3}(c) shows the measured Ramsey fringes after inserting 0, 100, 200, and 300 transfer operations into a constant total free evolution time. The fringe contrast and oscillation phase exhibit only minor changes even after 300 transfers, indicating that the destruction of qubit coherence by the transfer process is negligible.

\cref{fig:s3}(d) shows the Ramsey fringes measured after inserting 0, 20, 60, and 100 transport operations into a constant total evolution time. Although fewer cycles can be performed due to the longer operation time, the fringe contrast remains excellent.

\end{document}